\documentclass[prd,preprint,superscriptaddress,amsmath,amssymb,nofootinbib]{revtex4}
\pdfoutput=1

\usepackage{graphicx,color,slashed}
\usepackage{subfigure}
\def\be{\begin{eqnarray}}
\def\ed{\end{eqnarray}}

\begin{document}


\title{\bf \Large Light charged Higgs boson with dominant decay to a charm quark and a bottom quark 
and its search at LEP2 and future $e^+e^-$ colliders}

\author{A.G. Akeroyd}
\email{a.g.akeroyd@soton.ac.uk}
\affiliation{School of Physics and Astronomy, University of Southampton,
Highfield, Southampton SO17 1BJ, United Kingdom}

\author{Stefano Moretti}
\email{S.Moretti@soton.ac.uk}

\affiliation{School of Physics and Astronomy, University of Southampton,
Highfield, Southampton SO17 1BJ, United Kingdom}

\author{Muyuan Song}
\email{ms32g13@soton.ac.uk}

\affiliation{School of Physics and Astronomy, University of Southampton,
Highfield, Southampton SO17 1BJ, United Kingdom}

\date{\today}

\begin{abstract}
The possibility of a light charged Higgs boson $H^\pm$ that decays 
predominantly to $cb$ and with a mass in the
range 80 GeV $\le M_{H^\pm} \le 90$ GeV
is studied in the context of a 3-Higgs Doublet Model (3HDM). 
Searches for this decay at the Large Hadron Collider (LHC) do not have sensitivity to this mass region at present.
It is shown that the searches for $H^\pm$ at LEP2 could be supplemented by
either one or two $b$-tags, which would enable such large branching ratios for $H^\pm\to cb$ 
to be probed in the above mass region. We comment on the 
possibility of this 3HDM scenario to explain a slight excess in the searches 
for $H^\pm$ at LEP2, which
is best fit by $M_{H^\pm}$ of around 90 GeV, and discuss the prospects for 
detecting $H^\pm \to cb$ decays at future $e^+e^-$ colliders.

\end{abstract}

\maketitle

\section{Introduction}
\noindent
The ATLAS and CMS \cite{Aad:2012tfa,Chatrchyan:2012xdj} collaborations at the CERN Large Hadron Collider (LHC)
announced the discovery of a new particle (a spinless boson) with a mass of 125 GeV. The measurements of its properties (couplings, spin etc)
are in excellent agreement with those of the Higgs boson of the Standard Model (SM), in 
which the Higgs boson originates from an $SU(2)\otimes U(1)$ scalar doublet.

It is possible that the 125 GeV boson is the first scalar to be discovered from a non-minimal Higgs sector.
A (singly) electrically charged Higgs boson $H^\pm$ would represent a distinctive signal of 
such a structure (see Ref.~\cite{Akeroyd:2016ymd} for a recent phenomenological review) that could include
additional doublets, singlets, triplets or combinations thereof.  
There is considerable interest in Beyond the SM (BSM) scenarios with such a framework for
implementing the Higgs mechanism of Electro-Weak Symmetry Breaking (EWSM). 
Firstly, the SM is non-minimal in both its matter (with three fermionic generations) and gauge (with both strong and EW 
force mediators) sectors, and so there is no compelling reason to believe that the Higgs sector should be minimal. 
Secondly, in some BSM scenarios an enlarged Higgs sector is required theoretically (e.g. supersymmetry)
or provides an explanation to problems that are not solved in the SM
 (e.g. necessity of non-zero neutrino masses, requirement of a dark matter candidate, sufficient EW baryogenesis etc). 

The 2-Higgs Doublet Model (2HDM) \cite{Gunion:1989we,Branco:2011iw} has attracted the most attention among 
models with additional scalar doublets. Two (softly-broken) discrete $Z_2$ symmetries are imposed in order to
ensure that each fermion-type couples to no more than one scalar doublet, leading to four distinct 
2HDMs that differ in their Yukawa couplings. This framework,
referred to as ``Natural Flavour Conservation'' (NFC) \cite{Glashow:1976nt}, is invoked 
in order to avoid Flavour Changing Neutral Currents (FCNCs) that are mediated at tree-level by neutral scalars.
More recently,  3-Higgs Doublet Models (3HDMs) have received increased attention 
(see, e.g. Refs.  \cite{Ivanov:2012fp,Keus:2013hya} for mini-reviews), with NFC leading to five
distinct 3HDMs.

Regarding the particle content of the 3HDM there are two physical charged Higgs bosons 
(hereafter denoted by $H^\pm$ and $H^{'\pm}$, with $M_{H^\pm}<M_{H^{'\pm}}$). 
More parameters determine the phenomenology of the charged Higgs sector than in 2HDMs,
and we make the assumption that all three Higgs doublets have a 
Vacuum Expectation Value (VEV).
In Refs.~\cite{Grossman:1994jb, Akeroyd:1994ga, Cree:2011uy, Akeroyd:2012yg,Akeroyd:2016ssd}, 
the phenomenology of $H^\pm$ in 3HDMs has been studied (with decoupled $H^{'\pm}$) in terms of 
effective Yukawa couplings for the down-type quark, 
up-type quark and charged lepton, 
which are expressed as a function of four independent parameters \cite{Cree:2011uy} in the framework of NFC. 
It has been shown that a  
$H^\pm$ can be lighter than the top quark (with $H^{'\pm}$ heavier)
while satisfying constraints from $B\to X_s\gamma$ (even for the Yukawa coupling combinations 
that would not permit this scenario in the 2HDM) due to the increased number of parameters in the
3HDM and the presence of two charged scalars. 
Moreover, it was shown in Refs. \cite{Grossman:1994jb, Akeroyd:1994ga,Akeroyd:1998dt,Akeroyd:2012yg} that the decay channel 
$H^+\to c\bar b$ can have a large Branching Ratio (BR)
(up to $80\%$) in a 3HDM. Although such a value for this BR is theoretically allowed in the flipped 2HDM for 
$M_{H^\pm}< m_t-m_b$ \cite{Akeroyd:1994ga},
the constraint $M_{H^\pm}> 480$ GeV from $B\to X_s\gamma$ rules out this possibility. Hence a large BR($H^+\to c\bar b$) is 
a distinctive signature of 3HDMs.

The above scenario of a 3HDM in which there is a light $H^\pm$ with a large BR to $c\bar b$ is the focus of this work.
We consider the mass range $M_{H^\pm}\sim M_{W^\pm}$ for which detection of $H^\pm$ is challenging if its 
BRs to hadrons are dominant.
The LHC has carried out searches at $\sqrt s=8$ TeV for $H^\pm\to cb$ \cite{Sirunyan:2018dvm} and $H^\pm\to hadrons$ 
\cite{Aad:2013hla,Khachatryan:2015uua}, assuming production
via the mechanism $t\to H^\pm b$, and the former search employs one more $b$-tag than the latter search.
In Ref. \cite{Akeroyd:2018axd} the parameter space in the flipped 
3HDM that will be excluded (or provide a signal)
at upcoming searches was displayed. At present, the LHC
 has not set limits in the region $80$ GeV $\le M_{H^\pm} \le 90$ GeV if BR$(H^+ \to c\bar b$) {\rm or} BR$(H^+\to c\bar s)$ is dominant, 
although limits are set for the case of $H^\pm\to \tau\nu$ being the leading decay channel. As discussed in \cite{Akeroyd:2018axd},
with the increased luminosity for the data taken at $\sqrt s=13$ TeV and with future data 
it is likely that the LHC will be able to set limits on BR$(t\to H^+ b)$ $\times$ BR$(H^+\to c\bar b)$ in part (if not all) of the region 
$80$ GeV $ \le M_{H^\pm} \le 90$ GeV (and possibly in the case of the $H^+\to c\bar s$ channel as well).
However, the production mechanism relies on the Yukawa couplings and thus such an $H^\pm$ could escape detection at the LHC
if these couplings are small. Consequently, it is
of interest to study in more detail the CERN LEP2 searches for a hadronically decaying $H^\pm$, for which the main production
mode of $e^+e^-\to H^+H^-$ depends only on gauge couplings and $M_{H^\pm}$.

We will show that data taken at LEP2 when supplemented by $b$-tagging could discover or exclude a light $H^\pm$ state 
decaying to $c\bar b$ pairs  
more efficiently than LHC searches in the region $80$ GeV $ \le M_{H^\pm} \le 90$ GeV.
Before the LEP2 era this possibility was pointed out for models with more than two Higgs doublets 
in Refs. \cite{Akeroyd:1994ga,Akeroyd:1998dt}, although the brief quantitative study in \cite{Akeroyd:1998dt} 
(that was based on a simulation in \cite{Sopczak:1993jt}) concluded that
sensitivity would not be reached in the region $80$ GeV $ \le M_{H^\pm} \le 90$ GeV.
Such a $b$-tag was never implemented in LEP2 searches for $H^\pm$ states. 
We revisit it here in the context of the flipped 3HDM and 
show that by using $b$-quark tagging and light-quark rejection efficiencies from the LEP2 searches one can substantially improve the sensitivity 
to $H^\pm\to cb$ compared to that for $H^\pm\to hadrons$, and probe the region $80$ GeV $ \le M_{H^\pm} \le 90$ GeV.
Attention is also given to the detection prospects
for $H^\pm \to cb$ at future $e^+e^-$ colliders operating at $\sqrt s=240$ GeV.

The plan of this paper is as follows. In section II the 3HDM is introduced. In section III the LEP2 search for $H^\pm$
with the addition of $b$-tagging is described, with numerical results and conclusions in sections IV and V respectively.

\section{THE 3HDM WITH NFC}

In this section we give a brief introduction to the interactions of the 
lightest $H^\pm$ in the 3HDM 
that are relevant to our analysis. We will only consider $M_{H^\pm}<m_t$ and we assume that the only channels
that have non-zero BRs are the decays to fermions (i.e. decays of the type 
$H^\pm\to W^\pm$ plus a neutral Higgs 
boson are forbidden
by setting the masses of all the neutral Higgs bosons to be above that of the charged Higgs). 
For a more detailed introduction the reader is referred to \cite{Cree:2011uy,Akeroyd:2016ssd}.

Any extension of the SM Higgs sector is primarily constrained by two experimental facts.
Firstly, the measurement of $\rho = m^2_W / (m^2_Z \cos^2\theta_W)$ is close to 1 \cite{Gunion:1989we}, 
where $m_W, m_Z$ and $\theta_W$ are the $W,Z$ masses and weak mixing angle, respectively. 
Secondly, tree-level FCNCs that are mediated by the additional neutral scalars must be suppressed (or absent). In order 
for the 
3HDM to comply with both of the above restrictions one requires i)   
no very large mass splittings between the neutral and charged scalars in order to respect 
$\rho$ parameter bounds, and ii) to implement NFC \cite{Glashow:1976nt}
in order to eliminate tree-level FCNCs.
Under such conditions, the part of the Yukawa Lagrangian containing the 
lightest charged Higgs boson interactions with the fermions can be written as follows:
\begin{equation}
\label{eqn:2la}
{ {\cal L}_{H^\pm} =
- H^{+} \{ \frac{\sqrt{2} V_{ud}}{v_{sm}} \bar{u} (m_d XP_R + m_u YP_L)d + }
\frac{\sqrt{2} m_\ell}{v_{sm}} Z \bar{\nu}_L \ell_R \} 
+{H.c.}
\end{equation}
Here $u (d)$ is denotes up(down)-type quarks and $\ell$ represents charged leptons, $P_{L(R)}$ is the Left(Right)-handed projector, $V_{ud}$ is the relevant Cabibbo-Kobayashi-Maskawa (CKM) matrix element, and $v_{SM}$ is the VEV of the
Higgs doublet in the SM.
In the 3HDM, the couplings $X,Y$ and $Z$ are functions of the four parameters (see below) of a unitary
matrix $U$ that connects the charged scalar interaction eigenstates to the physical mass eigenstates as follows:
\begin{equation}
\left( \begin{array}{c} G^+ \\ H^+ \\ H^{'+} \end{array} \right) 
= U \left( \begin{array}{c} \phi_1^+ \\ \phi_2^+ \\ \phi_3^+ \end{array} \right).
\end{equation}
Here $H^{+},H^{'+}$ are physical charged scalars whereas  $G^+$ is a charged Goldstone boson that will 
become the longitudinal component of the $W^{\pm}$ gauge boson after EWSB. 
The matrix $U$ is a 3 $\times$ 3 unitary matrix and can be parametrised as a function of four parameters, 
$\tan\beta, \tan\gamma,\theta,$ and $\delta$. The first two parameters are defined via 
\begin{equation}
\tan\beta = v_2/v_1,  \quad
\tan\gamma = \sqrt{v^2_1 + v^2_2}/v_{3},
\end{equation}
where $v_1,v_2,$ and $v_3$ are the VEVs of each Higgs doublet. The parameter 
$\theta$ is a mixing angle between the two massive charged scalars
  and $\delta$ is a CP-violating phase. The explicit form of $U$
is as follows \cite{Cree:2011uy}:
\begin{equation}\label{eq:Uexplicit}
U =  \left( \begin{array}{ccc}
	s_\gamma c_\beta & s_\gamma s_\beta 	& c_\gamma \\
	-c_\theta s_\beta e^{-i\delta} - s_\theta c_\gamma c_\beta 
		& c_\theta c_\beta e^{-i\delta} - s_\theta c_\gamma s_\beta & s_\theta s_\gamma \\
	s_\theta s_\beta e^{-i\delta} - c_\theta c_\gamma c_\beta 
		& -s_\theta c_\beta e^{-i\delta} - c_\theta c_\gamma s_\beta & c_\theta s_\gamma 
	\end{array} \right),
\end{equation}	
where $s(c)$ are represents the sine(cosine) of the respective angle.

The interactions  between the lightest charged Higgs state of the 3HDM, $H^{\pm}$, and the SM fermions is obtained via the $U$ matrix  as \cite{Grossman:1994jb}
\begin{equation}\label{eq:X_Y_Z}
X = \frac{U_{d2}^\dagger}{U_{d1}^\dagger}, \quad \quad 
Y= - \frac{U_{u2}^\dagger}{U_{u1}^\dagger}, \quad \quad 
Z = \frac{U_{\ell 2}^\dagger}{U_{\ell1}^\dagger},
\end{equation}\\
where the values of $d$, $u$, and $\ell$ in these matrix elements are given in Tab. \ref{tab:3HDMNFC} and depend upon
which of  the five possible distinct 3HDMs is under consideration. Taking $d = 1, u = 2$ and $\ell = 3$
means that the down-type quarks receive their mass from
$v_1$, the up-type quarks from $v_2$ and the charged leptons from $v_3$. This choice is called the
`Democratic 3HDM' while the other possible choices of $d, u$ and $\ell$ in a 3HDM are given the
same names as the four standard types of 2HDM \cite{Branco:2011iw}. 

\begin{table} [!t]
\begin{center}
\begin{tabular}{|c||c|c|c|}
\hline
& { $u$} & { $d$} &  { $\ell$} \\ \hline
3HDM(Type I)
&  {\color{black} $2$} & {\color{black} $2$} & {\color{black} $2$} \\
3HDM(Type II)
& {\color{black} $2$} & {\color{black} $1$} & {\color{black} $1$} \\
3HDM(Lepton-specific)
& {\color{black} $2$} & {\color{black} $2$} & {\color{black} $1$} \\
3HDM(Flipped)
& {\color{black} $2$} & {\color{black} $1$} & {\color{black} $2$} \\
3HDM(Democratic)
& {\color{black} $2$} & {\color{black} $1$} & {\color{black} $3$} \\
\hline
\end{tabular}
\end{center}
\caption{\label{tab:3HDMNFC} The five versions of the 3HDM with NFC and the corresponding $u,d$ and $\ell$ values. Taking $u = i$ means that the up-type quarks receive their mass from $v_i$ and likewise for $d$ (down-type quarks) and $\ell$ (charged leptons). }
\end{table}

The experimental constraints on $X,Y$ and $Z$
\cite{Jung:2010ik,Trott:2010iz} have been 
summarised in Ref. \cite{Akeroyd:2018axd}, to which we refer the reader.
The parameter space of the 3HDM that is relevant to this
work is compliant with all such limits,
the most important of which being $-1.1 < {\rm Re}(XY^*)< 0.7 $
for $M_{H^\pm}<100$ GeV. This is an approximate constraint that is derived
from $b\to s\gamma$, and neglects the contribution of the heavier $H^{\prime\pm}$
in a 3HDM.

In a 3HDM, the expressions for the partial widths of the decay of $H^\pm$ to fermions
are as follows:

\begin{equation}
\Gamma(H^\pm\to \ell^\pm\nu)=\frac{G_F M_{H^\pm} m^2_\ell |Z|^2}{4\pi\sqrt 2}\;, 
\label{width_tau}
\end{equation}
\begin{equation}
\Gamma(H^\pm\to ud)=\frac{3G_F V_{ud}M_{H^\pm}(m_d^2|X|^2+m_u^2|Y|^2)}{4\pi\sqrt 2}\;.
\label{width_ud}
\end{equation}

In the expression for 
$\Gamma(H^\pm\to ud)$ the running quark masses should be evaluated at the 
scale of $m_{H^\pm}$, and
there are QCD vertex corrections which multiply the partial widths by
$(1+17\alpha_s/(3\pi))$. 
The first study of the fermionic BRs of $H^\pm$ as a function of $|X|$, $|Y|$, and $|Z|$ 
was given in \cite{Akeroyd:1994ga}, with further studies in \cite{Akeroyd:2012yg}.
In \cite{Akeroyd:2016ssd,Akeroyd:2018axd} these BRs were studied as a function of 
 $\tan\beta, \tan\gamma, \theta,$ and $\delta$, an approach which allows the BRs in the five versions of the 3HDM
to be compared.
For $|X|\gg |Y|,|Z|$ the decay channel BR$(H^\pm\to cb)$ dominates
(which was first mentioned in \cite{Grossman:1994jb}), and 
reaches a maximum of $\sim 80\%$. It was shown in \cite{Akeroyd:2016ssd,Akeroyd:2018axd} 
that such large values of BR$(H^\pm\to cb)$ are only possible in the flipped and democratic 3HDMs,
with  BR$(H^\pm\to cb)$ having a maximum value of around 1\% in the other 3HDMs.  
In 2HDMs with NFC the only model which contains a 
parameter space for a large BR$(H^\pm\to cb)$ with $M_{H^\pm}< m_t$
is the flipped model (a possibility that was mentioned in 
\cite{Grossman:1994jb,Akeroyd:1994ga} and studied in more detail in 
\cite{Logan:2010ag}).
However, for this particular choice of 2HDM the $b\to s\gamma$ constraint 
would require
$M_{H^\pm}>500$ GeV \cite{Hermann:2012fc,Misiak:2015xwa} for which
$H^\pm\to tb$ would dominate.

\begin{table}[h]
\begin{center}
\begin{tabular}{|c||c|c|}
\hline
& ATLAS &  CMS  \\ \hline
7 TeV (5 fb$^{-1}$)
&  $cs$ \cite{Aad:2013hla}, $\tau\nu$ \cite{Aad:2012rjx,Aad:2012tj}
&  $\tau\nu$ \cite{Chatrchyan:2012vca} \\
8 TeV (20 fb$^{-1}$)
& $\tau\nu$ \cite{Aad:2014kga} & $cs$ \cite{Khachatryan:2015uua}, 
$cb$ \cite{Sirunyan:2018dvm}, 
$\tau\nu$ \cite{Khachatryan:2015qxa}  \\
13 TeV (36 fb$^{-1}$)
& $\tau\nu$ \cite{Aaboud:2018gjj} & $\tau\nu$  \cite{Sirunyan:2019hkq} \\
\hline
\end{tabular}
\end{center}
\caption{Searches for $H^\pm$ at the LHC, using $pp\to t\overline t$ and $t\to H^\pm b$. The given integrated
luminosities are approximate. The search in \cite{Chatrchyan:2012vca}
used 2 fb$^{-1}$.}
\label{LHC_search}
\end{table}

In this paper we will focus on the case of $m_{H^\pm}< m_t$, a scenario in which production at the LHC via $t\to H^\pm b$ would
be possible. Searches for three decays channels of $H^\pm$ have been carried out (see Tab.~\ref{LHC_search}). 
The searches for $H^\pm \to \tau\nu$ constrain the product BR$(t\to H^\pm b)\times {\rm BR}(H^\pm\to \tau\nu)$ in the
region 80 GeV$< M_{H^\pm} < 160$ GeV, with the upper limit ranging from 
$<0.36\%$ for $M_{H^\pm}=80$ GeV to  $<0.08\%$ for $M_{H^\pm}=160$ GeV.
The searches for $H^\pm \to cs$ constrain the product BR$(t\to H^\pm b)\times {\rm BR}(H^\pm\to cs)$ in the
region 90 GeV$< M_{H^\pm} < 160$ GeV, with the upper limit ranging from 
$<5\%$ for $M_{H^\pm}=90$ GeV to  $<2\%$ for $M_{H^\pm}=160$ GeV. Note that this search would be sensitive to
any quark decay (except $t$) of $H^\pm$. The search for $H^\pm \to cb$ (which employs one more $b$-tag than the
search for $H^\pm\to cs$) constrains the product BR$(t\to H^\pm b)\times {\rm BR}(H^\pm\to cb)$, 
with  the upper limit ranging from $<1.4\%$ for $M_{H^\pm}=90$ GeV to  
$<0.5\%$ for $M_{H^\pm}=150$ GeV.
The searches for $H^\pm\to cs$ and  $H^\pm\to cb$ do not set limits on the region 80 GeV$< M_{H^\pm} < 90$ GeV,
although this might be possible (especially for $H^\pm\to cb)$ with larger integrated luminosities. Earlier
searches for the decay $t\to H^\pm b$ were carried out at the Fermilab Tevatron in \cite{Abazov:2009aa,Aaltonen:2009ke}.

At LEP2 the production process $\sigma(e^+ e^- \to \gamma^*, Z^* \to H^{+} H^{-} )$ was used, which depends on only one unknown parameter, $M_{H^{\pm}}$.
Searches were carried out at all four experiments  
\cite{Abbiendi:2008aa,Heister:2002ev,Achard:2003gt,Abdallah:2003wd}
at energies in the range $\sqrt s=183$
GeV  to $\sqrt s=209$ GeV, each with an integrated luminosity
of roughly 0.6 fb$^{-1}$. The LEP working group  \cite{Abbiendi:2013hk} combined 
these individual searches, resulting in a cumulative integrated luminosity
of 2.6 fb$^{-1}$.
Dedicated searches for the decay mode $H^\pm\to A^0W^*$ were also carried 
out in \cite{Abbiendi:2008aa,Abdallah:2003wd}, but in this work we are assuming that
this channel is absent or very suppressed.
From the combination of the 
searches for fermionic decays, 
and with the assumption that
BR$(H^\pm\to \tau\nu$)+BR$(H^\pm\to cs$)=1,
the excluded region at 95\% Confidence Level  
in the plane $[M_{H^\pm}, {\rm BR}(H^\pm\to \tau\nu)]$
is obtained in  \cite{Abbiendi:2013hk}.
For $M_{H^\pm}<80$ GeV the whole range 
$0 \le {\rm BR}(H^\pm\to \tau\nu)\le 100\%$ is excluded. For $80\; {\rm GeV}\le \;M_{H^\pm}<90$ GeV, most of the region is not excluded
for BR$(H^\pm \to \tau\nu)<80\%$ (i.e.  for BR$(H^\pm \to cs)>20\%$).
We will focus on this region of $80\; {\rm GeV}\le \;M_{H^\pm}<90$ GeV
and the case of a large hadronic BR for $H^\pm$, 
which is not being probed by the LHC at present.

\section{Search for $H^\pm$ at LEP2}\label{LEP}

At LEP2 it was assumed that the dominant decay channels were $H^{\pm} \to cs$ and $H^{\pm} \to \tau\nu$, which leads to the 
following three signatures from $H^+H^-$ production: $cscs, cs\tau\nu, \tau\nu\tau\nu$. The decay of $H^{\pm} \to cb$ was not explicitly searched for at LEP2 \cite{Abbiendi:2008aa,Abdallah:2003wd,Heister:2002ev,Achard:2003gt}.
 It is the searches in the hadronic channels $cscs$ and $cs\tau\nu$ that are relevant for the decay $H^{\pm} \to cb$, and these are discussed in more detail below.

i)   4-jet channel: This signature arises when $H^+$ and $H^-$ both decay into quarks, 
giving four quarks that will usually be detected as 4 jets. For $H^{\pm}$ in the kinematical range of LEP2 (i.e. $M_{H^{\pm}} < \sqrt{s} / 2 \approx 100$ GeV) there are six possible hadronic decay channels of $H^{\pm}$. Decays involving the $t$ quark (e.g. $H^{\pm} \to t^* b$) are extremely suppressed 
due to the t quark being very off-shell, and can be neglected. In the LEP searches it was assumed that $H^{\pm} \to cs$ is the dominant hadronic decay mode, which is true in most 2HDMs, and the experimental limits on BR($H^{\pm} \to hadrons$) were interpreted as limits on BR($H^{\pm} \to cs$).  However, the 4-jet search as carried out by three of the LEP collaborations 
(OPAL \cite{Abbiendi:2008aa}, ALEPH \cite{Heister:2002ev}, L3 \cite{Achard:2003gt}) was sensitive to any of the allowed six decay channels into quarks. In contrast, the search by the DELPHI collaboration \cite{Abdallah:2003wd}
used $c$-tagging to discriminate against lighter quarks and $b$ quarks. Consequently, this search strategy would be less sensitive to the decay $H^{\pm} \to cb$ than the searches by the other three collaborations.

ii)  2-jet$+\tau\nu$ channel: This signature arises when one $H^{\pm}$ decays into quarks and the other $H^{\pm}$ decays into a $\tau$ lepton and a neutrino. Again, it was assumed that  $H^{\pm} \to cs$ is the dominant hadronic decay mode, and the DELPHI collaboration alone used $c$-tagging.\\

In this work we quantify the effect of applying one (or more) $b$-tags to both of the above
search strategies in order to increase the sensitivity to the decay $H^{\pm} \to cb$, which can have
a large BR in the flipped and democratic 3HDMs. In the 4-jet channel the
separate cases of exactly one tagged $b$-jet and exactly two tagged $b$-jets will be considered.
In the 2-jet$+\tau\nu$ channel the case of exactly one tagged $b$-jet will be considered. A $b$-tag requirement usually involves a cut on the impact parameter of a jet \cite{Abdallah:2002xm}.
Due to the longer lifetime of the $b$ quark, a jet that has originated from a $b$ quark will (on average) have a
larger impact parameter than a jet that originated from a lighter quark. Additional discriminating variables are sometimes used in the full $b$-tag requirement. The three dominant
decay channels of $H^{\pm}$ in the 3HDMs that we study are BR($H^{\pm} \to cb$), BR($H^{\pm} \to cs$) and
BR($H^{\pm} \to \tau\nu$). These will be denoted below by $BR_{cb} $, $BR_{cs} $, and $BR_{\tau \nu}$ respectively.

\subsection{ Signal for $H^{\pm} \to cb$ with $b$-tags at LEP2}
The number of $e^{+} e^- \to H^+ H^-$ events (with no $b$-tag requirement) in the LEP2 searches in the 4-jet and 2-jet$+\tau\nu$ channels are denoted by $S_{4jnobtag}$ and $S_{2j \tau nobtag}$ respectively, and
are given as follows:

(i)$S_{4jnobtag}=\sigma\times{\cal L}\times\epsilon_{4jnobtag}\times(BR_{cb}+BR_{cs})^2$. Note that $BR_{cb}$ and $BR_{cs}$ are summed, because the search strategy does not apply a $b$-tag.

(ii)$S_{2j{+\tau}nobtag}=\sigma\times{\cal L}\times\epsilon_{2j{\tau}nobtag} \times 2(BR_{cb}+BR_{cs})BR_{\tau\nu}$. Note that $BR_{cb}$ and $BR_{cs}$ are summed (as above), and the factor of 2 accounts for the separate contributions from $c \bar{s} \tau^- \bar{\nu}$ and $\bar{c} s \tau^+ \nu$.\\
Here $\sigma$ is the cross-section for pair production of $H^+ H^-$ at a particular centre-of-mass energy $\sqrt{s}$, and $\cal L$ is integrated luminosity at that energy. The searches for $H^+ H^-$ at LEP2 were carried out using
data taken at eight different values of $\sqrt{s}$, each with a unique value of integrated luminosity $\cal L$. Hence the product $\sigma \cal L$ is actually a sum $\sum^8_{i = 1} \sigma_i {\cal L}_i $ where each $i$ denotes a specific value of $\sqrt{s}$. The parameters  $\epsilon_{4jnobtag}$ and $\epsilon_{2j \tau nobtag}$ are the selection efficiencies for the cuts as used in the LEP searches for the 4-jet signature and the 2-jet$+\tau\nu$ signature respectively. For the magnitude of these efficiencies we will use the numerical values obtained in the search by OPAL (similar values were obtained by the other three collaborations). We now discuss in turn three proposed search strategies for the decay $H^{\pm} \to cb$ that make use of $b$-tagging.\\

\newpage
\textit{1. Signal in 4-jet channel with exactly two $b$-tagged jets}\\

A maximum of two $b$ quarks can be produced when both charged scalars decay via $H^{\pm} \to cb$. However, lighter quarks ($u,d,s,c$) can fake $b$ quarks, and so up to four jets could be recorded as $b$-jets by a detector. In the numerical analysis for LEP2 the $b$-tag efficiency ($\epsilon_b$) is taken to be $\epsilon_b$ = 0.7, while the fake $b$-tag efficiencies for charm quarks ($\epsilon_c$) and $u,d,s$ quarks ($\epsilon_j$) are $\epsilon_c$ = 0.06 and $\epsilon_j$= 0.01 respectively. These numbers
are roughly similar (although slightly optimistic for $\epsilon_b$) to those in the OPAL measurement of $R_b$ in \cite{Abbiendi:2004vw}
for $\sqrt s=183\,{\rm GeV}$ to $209\,{\rm GeV}$.
Due to $\epsilon_c$ and $\epsilon_j$ being small we will not consider the signatures of three or four tagged $b$-jets, in which one or two non-$b$ quarks have been mistagged as $b$ quarks. We first consider the channel in which exactly two of the four jets are tagged as $b$ jets. The number of such events is denoted by $S_{4j2btag}$, and is given by the following expression:
\begin{equation}
S_{4j2btag}=\sigma\times{\cal L}\times\epsilon_{4jnobtag}\times(BR_{cb}BR_{cb}\epsilon^{cbcb}_{4j2btag}+2BR_{cb}BR_{cs}\epsilon^{cbcs}_{4j2btag}+BR_{cs}BR_{cs}\epsilon^{cscs}_{4j2btag})\,.
\end{equation}
The factor of 2 accounts for the $c \bar{b} \bar{c}s$ and $\bar{c}bc\bar{s}$ signatures. This expression for $S_{4j2btag}$ is obtained from the expression for $S_{4jnobtag}$, with the effect of the $b$-tagging requirement contained in the parameters $\epsilon^{cbcb}_{4j2btag}$,  $\epsilon^{cbcs}_{4j2btag}$ and $\epsilon^{cscs}_{4j2btag}$ that are given explicitly as follows:
\begin{equation}
\begin{aligned}
\epsilon^{cbcb}_{4j2btag} &= \epsilon^{2}_{b} (1-\epsilon_c)^2+ 4\epsilon_b\epsilon_c(1-\epsilon_b)(1-\epsilon_c)+\epsilon_c^2(1-\epsilon_b)^2\,,\\
\epsilon^{cbcs}_{4j2btag} &= \epsilon_b\epsilon_c(1-\epsilon_c)(1-\epsilon_l)+\epsilon_b\epsilon_l(1-\epsilon_c)^2 + 2\epsilon_c\epsilon_l(1-\epsilon_b)(1-\epsilon_c) + \epsilon_c^2(1-\epsilon_b)(1-\epsilon_l)\,,\\
\epsilon^{cscs}_{4j2btag}  &= 4\epsilon_c\epsilon_l(1-\epsilon_c)(1-\epsilon_l)+\epsilon_c^2(1-\epsilon_l)^2 + \epsilon_l^2(1-\epsilon_c)^2\,.
\end{aligned}
\end{equation}
Inserting the above values for  $\epsilon_b$, $\epsilon_c$ and $\epsilon_j$ gives numerical values of roughly 0.48, 0.04 and 0.004 for $\epsilon^{cbcb}_{4j2btag}$,  $\epsilon^{cbcs}_{4j2btag}$ and $\epsilon^{cscs}_{4j2btag}$ respectively. Note that the three terms in $\epsilon^{cbcb}_{4j2btag}$
correspond to the cases of the two tagged $b$-jets originating from i) two real $b$ quarks, ii) one real $b$ quark and one fake $b$ quark (i.e. a mistagged $c$ quark), and iii) two fake $b$ quarks. In $\epsilon^{cbcs}_{4j2btag}$ the first two terms correspond to the case of the two tagged $b$-jets originating from one real $b$ quark and one fake $b$ quark, and the last two terms are for the case of two fake $b$ quarks. In $\epsilon^{cscs}_{4j2btag}$ the only contributing terms are from two fake $b$ quarks. Factors of 2 or 4 in these expressions account for the various combinations that contribute (e.g. $c \bar{s}$ and $\bar{c} s$ being the fake $b$-tags in the third term in $\epsilon^{cbcs}_{4j2btag}$, leading to a factor of 2).\\

\textit{2. Signal in 4-jet channel with exactly one $b$-tagged jet}\\

The number of 4-jet events in which exactly one of the jets is tagged as a $b$ quark is denoted by $S_{4j1btag}$, and is given by the following expression:\\
\begin{equation}
S_{4j1btag}=\sigma\times{\cal L}\times\epsilon_{4jnobtag}\times(BR_{cb}BR_{cb}\epsilon^{cbcb}_{4j1btag}+2BR_{cb}BR_{cs}\epsilon^{cbcs}_{4j1btag}+BR_{cs}BR_{cs}\epsilon^{cscs}_{4j1btag})\,.
\end{equation}
The explicit expressions for $\epsilon^{cbcb}_{4j1btag}$, $\epsilon^{cbcs}_{4j1btag}$ and $\epsilon^{cscs}_{4j1btag}$ (which are different to those for the two $b$-tag case) are as follows:
\begin{equation}
\begin{aligned}
\epsilon^{cbcb}_{4j1btag}&= 2 \epsilon_b (1 - \epsilon_b) (1 - \epsilon_c)^2 + 2 (1 - \epsilon_b)^2 \epsilon_c 
(1 - \epsilon_c)\,,\\
\epsilon^{cbcs}_{4j1btag}&= \epsilon_b (1-\epsilon_c)^2 (1 - \epsilon_l) + 2 (1- \epsilon_b) \epsilon_c (1 - \epsilon_c) (1 - \epsilon_l) + \epsilon_l(1-\epsilon_b)(1-\epsilon_c)^2\,,\\
\epsilon^{cscs}_{4j1btag}&= 2 \epsilon_l (1 - \epsilon_c)^2 (1 - \epsilon_l) + 2 \epsilon_c (1 - \epsilon_c)(1- \epsilon_l)^2\,.
\end{aligned}
\end{equation}
Inserting the values for $\epsilon_b$, $\epsilon_c$ and $\epsilon_j$ gives numerical values of roughly 0.38, 0.64 and 0.13 for $\epsilon^{cbcb}_{4j1btag}$, $\epsilon^{cbcs}_{4j1btag}$ and $\epsilon^{cscs}_{4j1btag}$ respectively.\\

\textit{3. Signal in 2-jet plus $\tau \nu$ channel with exactly one $b$-tagged jet}\\

The number of 2-jet + $\tau \nu$ events in which exactly one of the jets is tagged as a $b$ quark is denoted by $S_{2j\tau 1btag}$, and is given by the following expression:\\
\begin{equation}
S_{2j\tau1btag}=\sigma\times{\cal L}\times\epsilon_{2j{\tau}nobtag} \times 2(BR_{cb}BR_{\tau\nu}\epsilon^{cb\tau\nu}_{2j{ \tau}1btag}+BR_{cs}BR_{\tau\nu}\epsilon^{cs\tau\nu}_{2j{\tau}1btag} )\,.
\end{equation}
The explicit expressions for $\epsilon^{cb\tau\nu}_{2j{\tau}1btag}$ and $\epsilon^{cs\tau\nu}_{2j{\tau}1btag}$ are as follows:
\begin{equation}
\begin{aligned}
\epsilon^{cb \tau\nu}_{2j{\tau}1btag}=
\epsilon_b ( 1 - \epsilon_c) + \epsilon_c (1 - \epsilon_b)\,,\\
\epsilon^{cs \tau\nu}_{2j{\tau}1btag}=
\epsilon_c (1 - \epsilon_l) + \epsilon_l (1 - \epsilon_c)\,.
\end{aligned}
\end{equation}
The numerical values of $\epsilon^{cb\tau\nu}_{2j{\tau}1btag}$ and $\epsilon^{cs\tau\nu}_{2j{\tau}1btag}$ are roughly 0.68 and 0.07 respectively.

\subsection{Background to $H^{\pm} \to cb$ decay}

The backgrounds for the above three channels are denoted by $B_{4j2btag}$, $B_{4j1btag}$ and $B_{2j{+\tau}1btag}$ respectively. The main contributions to $B_{4j2btag}$ and $B_{4j1btag}$ are from 4 fermion production (mainly $W^+W^-$ production, with a smaller contribution from $ZZ$) which we
neglect) and from 2-fermion production (e.g. $e^+ e^- \to \gamma^{*}, Z^{*} \to q\bar{q}gg$), which can give four jets. The main contribution to $B_{2j \tau  1btag}$ is from $W^+W^-$ production.

To evaluate the background before imposing $b$-tagging we again use the numbers in the OPAL search paper. For simplicity we assume a diagonal CKM matrix, and take BR($W^\pm \to cs$) = BR($W^\pm \to ud$) = 35\%. OPAL had around 1100 4-jet events after all cuts, of which 90\% are expected to be from 4-fermion events. With the assumption of a diagonal CKM matrix this background would be composed of 250 $cs\overline c \overline s$ events, 250 $ud\overline u \overline d$ events and 500 $csud$ events.
Given these numbers, it turns out that the contributions to the background from $W^\pm \to cb$ decays can be neglected because its branching ratio is about 600 times smaller than that of $W^\pm \to cs$. The contribution of 
$W^+W^- \to cb\overline c \overline b$ 
to the background would be much less than one event ($= 250 / 600^{2}$),  and the contributions from $W^+W^- \to cbcs$ and 
$W^+W^- \to cbud$ would each
be less than one event ($= 500/600$), before $b$-tagging is imposed.\\

\textit{1. Background to 4-jet channel with exactly two $b$-tagged jets}\\

The 4-fermion background to the 4-jet signal with two tagged $b$ quarks is given by:
\begin{equation}
B^{4fermion}_{4j2btag}=1000 \times (0.25\times\epsilon^{Wcscs}_{4j2btag}+0.5\times\epsilon^{Wcsud}_{4j2btag}+0.25\times
\epsilon^{Wudud}_{4j2btag})\,.
\end{equation}
The explicit expressions for $\epsilon^{Wcscs}_{4j2btag}$, $\epsilon^{Wcsud}_{4j2btag}$ and $\epsilon^{Wudud}_{4j2btag}$ are as follows:

\begin{equation}
\begin{aligned}
\epsilon^{Wcscs}_{4j2btag} &=
4 \epsilon_c \epsilon_l (1 - \epsilon_c) (1 - \epsilon_l) +\epsilon^2_c (1 - \epsilon_l)^2 + \epsilon^2_l(1 - \epsilon_c)^2\,,\\
\epsilon^{Wcsud}_{4j2btag}  &=  3\epsilon_c \epsilon_l (1 - \epsilon_l)^2  + 3\epsilon^2_l (1 - \epsilon_c)(1 - \epsilon_l)\,, \\
\epsilon^{Wudud}_{4j2btag} & =  4 \epsilon^2_l (1 - \epsilon_l)^2\,. 
\end{aligned}
\end{equation}
The numerical values of  $\epsilon^{Wcscs}_{4j2btag}$, $\epsilon^{Wcsud}_{4j2btag}$ and $\epsilon^{Wudud}_{4j2btag}$ are 0.006, 0.002 and 0.0004 respectively, giving $B^{4fermion}_{4j2btag} \approx 2$.

OPAL had around 100 4-jet events that originated from 2-fermion events. Around 15 of these would be $b\bar{b}$ events, due to $\sigma (e^+ e^- \to b\bar{b} )/ \sigma(e^+ e^- \to u\bar{u}, d\bar{d}, s\bar{s}, c\bar{c},b\bar{b})$ being roughly 0.15 at $\sqrt{s} = 200$ GeV. We estimate the 2-fermion background to the 4-jet signal with two tagged $b$ quarks to be:
\begin{equation}
B^{2fermion}_{4j2btag} = 15 \epsilon^2_b\,.
\end{equation}
This is around 7 events. The contribution to the 2-fermion background from $c\bar{c}$ events would be around 15$\epsilon^2_c$ and is much smaller than one event. The total background ($B_{4j2btag}$)  to the signal with 4-jets and two tagged $b$ quarks ($S_{4j2btag}$) is:
\begin{equation}
B_{4j2btag} = B^{4fermion}_{4j2btag} + B^{2fermion}_{4j2btag}\,.
\end{equation}
Since $B^{4fermion}_{4j2btag}$ is around 2 events, then the dominant background is from the 2-fermion events.\\

\textit{2. Background to 4-jet channel with exactly one $b$-tagged jet}\\

The 4-fermion background to the 4-jet signal with one tagged $b$-jet is given by:
\begin{equation}
B^{4fermion}_{4j1btag}=1000 \times (0.25\times\epsilon^{Wcscs}_{4j1btag}+0.5\times\epsilon^{Wcsud}_{4j1btag}+0.25\times\epsilon^{Wudud}_{4j1btag})\,.   
\end{equation}
The explicit expressions for $\epsilon^{Wcscs}_{4j1btag}$, $\epsilon^{Wcsud}_{4j1btag}$ and $\epsilon^{Wudud}_{4j1btag}$ are as follows:
\begin{equation}
\begin{aligned}
\epsilon^{Wcscs}_{4j1btag} & =
2 (1 - \epsilon_c )^2\epsilon_l(1 - \epsilon_l) + 2 \epsilon_c (1 - \epsilon_c) (1 - \epsilon_l )^2\,, \\
\epsilon^{Wcsud}_{4j1btag} & =  3 \epsilon_l(1- \epsilon_c)(1- \epsilon_l)^2  + \epsilon_c (1 - \epsilon_l)^3\,,\\
\epsilon^{Wudud}_{4j1btag} &= 4 \epsilon_l (1 - \epsilon_l)^3\,. 
\end{aligned}
\end{equation}
The numerical values of $\epsilon^{Wcscs}_{4j1btag}$, $\epsilon^{Wcsud}_{4j1btag}$ and $\epsilon^{Wudud}_{4j1btag}$ are 0.13,0.08 and 0.04 respectively.

We estimate the 2-fermion background (from $b\bar{b}$ production) to the 4-jet signal with one tagged b quark to be:
\begin{equation}
B^{2fermion}_{4j1btag} = 30 \epsilon_b (1 - \epsilon_b)\,.
\end{equation}
This is about 6 events, but is much less than the 4-fermion background, which is of the order of 90 events. We neglect the contribution to the 2-fermion background from $c\bar{c}$ events, which would be $30\epsilon_c (1 - \epsilon_c)$ and equal to around 1.7 events. Similar to before, one has:
\begin{equation}
B_{4j1btag} = B^{4fermion}_{4j1btag} + B^{2fermion}_{4j1btag}\,.
\end{equation}\\

\textit{3. Background to 2-jet plus $\tau\nu$ channel with exactly one $b$-tagged jet}\\

The background to the 2-jet plus $\tau\nu$ channel with exactly one $b$-tagged jet is dominantly from 4-fermion production, and is given by:
\begin{equation}
B^{4fermion}_{2j  {\tau}1btag}= 316 \times \frac{1}{2} \times (\epsilon^{Wcs\tau \nu}_{2j {\tau}1btag} + 
\epsilon^{Wud\tau \nu}_{2j {\tau}1btag})\,.
\end{equation}
The explicit expressions for $\epsilon^{Wcs\tau \nu}_{2j {\tau}1btag}$ and  $\epsilon^{Wud\tau \nu}_{2j {\tau}1btag}$ are as follows:
\begin{equation}
\begin{aligned}
\epsilon^{Wcs\tau \nu}_{2j {\tau}1btag} &=  \epsilon_c (1- \epsilon_l) + \epsilon_l (1 - \epsilon_c)\,.\\
\epsilon^{Wud\tau \nu}_{2j {\tau}1btag} &=  2 \epsilon_l (1 - \epsilon_l)\,. 
\end{aligned}
\end{equation}
The numerical values of $\epsilon^{Wcs\tau \nu}_{2j {\tau}1btag}$ and  $\epsilon^{Wud\tau \nu}_{2j {\tau}1btag}$ are 0.07 and 0.02 respectively.

\section{Numerical Results}
We now present our results for the statistical significances of a signal for $H^\pm\to cb$ at LEP2 ($189\,{\rm GeV}\le \sqrt s\le 209\,
{\rm GeV}$)
and at CEPC/FCC-ee ($\sqrt s=240$ GeV). In the context of LEP2 the region 80 GeV $\leq M_{H^{\pm}} \leq$ 90 GeV is studied, 
while at CEPC/FCC-ee we consider 80 GeV $\leq M_{H^{\pm}} \leq$ 120 GeV.
Of the five types of 3HDM the parameter space for a large BR$(H^\pm\to cb)$ is greatest in the flipped 3HDM, and hence results are shown in this model only. In our numerical analysis at CEPC/FCC-ee, $\epsilon_c$ is varied
in the range $0.01 < \epsilon_c < 0.06$, while $\epsilon_b$ and $\epsilon_j$ are conservatively taken to have the same values as at LEP2. 
Each LEP2 experiment accumulated around 0.6 fb$^{-1}$ of integrated luminosity ({$\cal L$}), while at CEPC/FCC-ee at least
1000 fb$^{-1}$ is expected. These input parameters are 
summarised in Tab.~\ref{tab:collider}:
\begin{table} [h]
\begin{center}
\begin{tabular}{|c||c|c|c|c|c|c|}
\hline 
& $\sqrt s$ & {$\cal L$}(fb$^{-1}$) &  $\epsilon_b$ &  $\epsilon_c$ &  $\epsilon_j$ &  $M_{H^\pm}$ \\ \hline
LEP2& 189 GeV $\to$ 209 GeV & 0.6 & 0.7 &  0.06 & 0.01 &  $80\,{\rm GeV}< M_{H^\pm} < 90$ GeV \\ \hline
CEPC/FCC-ee &  $240$ GeV &  1000 & 0.7  & $0.01 < \epsilon_c < 0.06$ & 0.01 &  $80\,{\rm GeV}< M_{H^\pm} < 120$ GeV \\
\hline
\end{tabular}
\end{center}
\caption{Input parameters used in the numerical analysis at LEP2 and at CEPC/FCC-ee.}\label{tab:collider}
\end{table}

\subsection{Enhancing the detection prospects for $H^\pm\to cb$ at LEP2 by using $b$-tags}
\begin{figure}[!b]
    \centering
     \begin{subfigure}
    {
        \includegraphics[scale=.5]{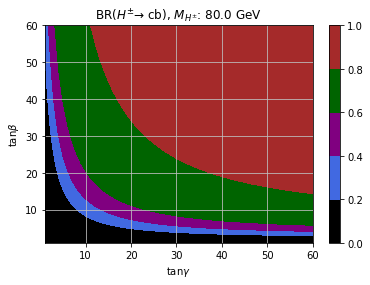}
    }
    {
        \includegraphics[scale=.5]{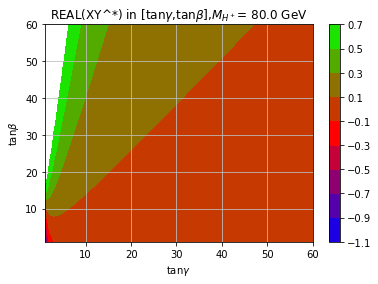}
    }
    \caption{The flipped 3HDM with $\theta = - \frac{\pi}{2.1}, \delta = 0$, and $M_{H^{\pm}} = 80$ GeV. Left Panel: Contours of 
BR$(H^\pm \to cb)$ in the plane $[\tan\gamma, \tan\beta$]. 
Right Panel: Contours of Re($XY^*$) where the region $-1.1 \le {\rm Re}(XY^*)\le 0.7$ is consistent with the limits on 
BR($b\to s\gamma$) for $M_{H^{\pm}} = 80$ GeV. }  \label{br80}
     \subfigure
    {
        \includegraphics[scale=.5]{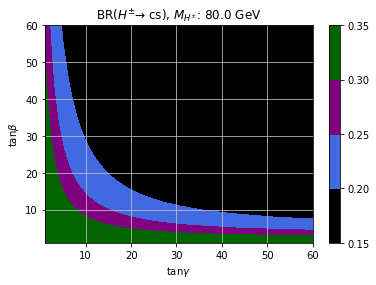}
    }
      \subfigure
    {
        \includegraphics[scale=.5]{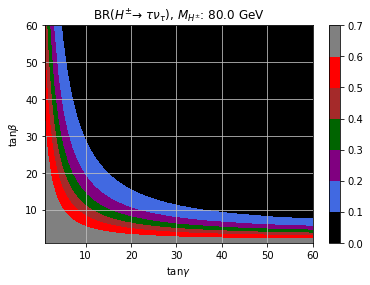}
    }
    \end{subfigure}
    \caption
    {The flipped 3HDM with $\theta = - \frac{\pi}{2.1}, \delta = 0$, and  $M_{H^{\pm}} = 80$ GeV. Left Panel: Contours of BR$(H^\pm \to cs)$ 
in the plane [$\tan\gamma, \tan\beta$].  Right Panel: Contours of BR$(H^\pm \to \tau\nu)$ in the plane [$\tan\gamma, \tan\beta$].
    } \label{brcbcs80}
\end{figure}

\begin{figure}[h!]
 	 \centering
    \subfigure
    {
        \includegraphics[scale=.55]{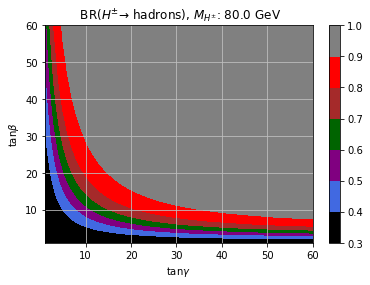}
    }
    \subfigure
    {
        \includegraphics[scale=.55]{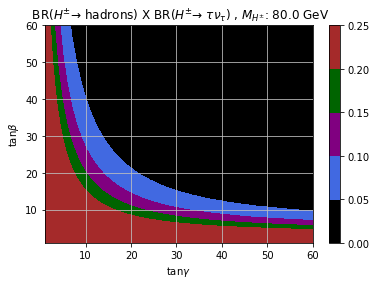}
    }\\
     \subfigure
    {
        \includegraphics[scale=.55]{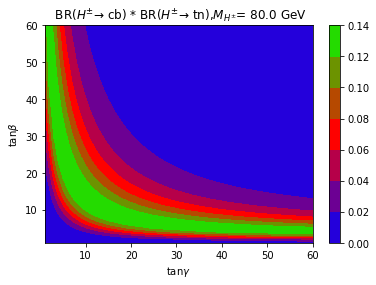}
    }
\caption{The flipped 3HDM with $\theta = - \frac{\pi}{2.1}, \delta = 0$, and  $M_{H^{\pm}} = 80$ GeV. Top Left Panel: Contours of 
BR$(H^\pm \to hadrons)$ in the plane [$\tan\gamma, \tan\beta$], where ``$hadrons$'' refers to the sum of $cs$ and $cb$.
        Top Right Panel: Contours of BR$(H^\pm \to hadrons) \times {\rm BR}(H^\pm \to \tau\nu)$ in the plane [$\tan\gamma, \tan\beta$].  
Bottom Panel: Contours of BR$(H^\pm \to cb) \times {\rm BR}(H^\pm \to \tau\nu)$ in the plane [$\tan\gamma, \tan\beta$]. }\label{background80}
\end{figure}

\begin{figure}[h!]
    \centering
    \subfigure
    {
        \includegraphics[scale=.5]{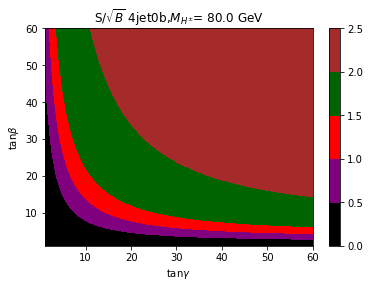}
    }\\
    \subfigure
    {
        \includegraphics[scale=.5]{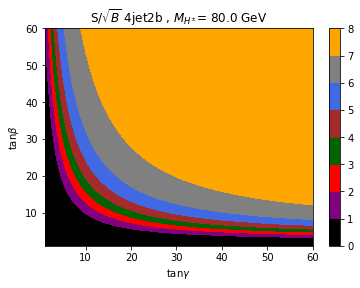}
    }
    \subfigure
    {
        \includegraphics[scale=.5]{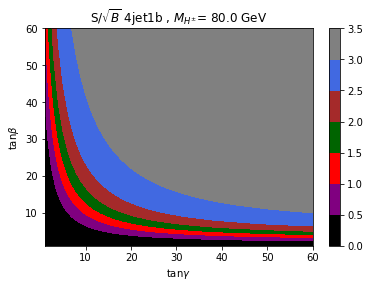}
    }
    \caption
    { The flipped 3HDM with $\theta = - \frac{\pi}{2.1}, \delta = 0$, and  $M_{H^{\pm}} = 80$ GeV. Top Panel: Significance ($S/\sqrt B$)
at a single LEP2 experiment in the 4-jet channel without $b$-tagging, in $[\tan\gamma, \tan\beta]$ plane. Left Bottom Panel: $S/\sqrt B$ 
in the 4-jet channel with two $b$-tags, in $[\tan\gamma, \tan\beta]$ plane. Right Bottom Panel: $S/\sqrt B$ in the 4-jet channel 
with one $b$-tag, in $[\tan\gamma, \tan\beta]$ plane.} \label{4j0b80}
\end{figure}

\begin{figure}[h!]
    \centering
    \subfigure[]
    {
        \includegraphics[scale=.55]{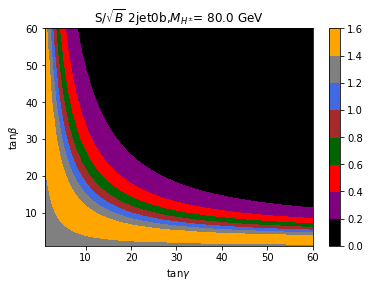}
    }
    \subfigure[]
    {
        \includegraphics[scale=.55]{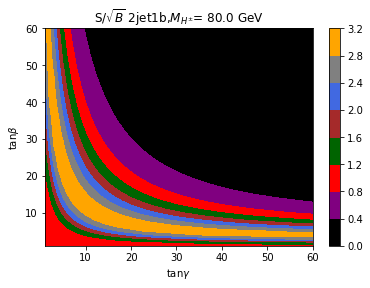}
    }
    \caption
    {The flipped 3HDM with $\theta = - \frac{\pi}{2.1}, \delta = 0$, and  $M_{H^{\pm}} = 80$ GeV. Left Panel: Significance ($S/\sqrt B$) 
at a single LEP2 experiment in the 2-jet channel without $b$-tagging, in $[\tan\gamma, \tan\beta]$ plane. 
Right Panel: $S/\sqrt B$ in the 2-jet channel with one $b$-tag, in $[\tan\gamma, \tan\beta]$ plane.
    }\label{2jallb80}
\end{figure}

\begin{figure}[h!]
    \centering
    \subfigure
    {
        \includegraphics[scale=.55]{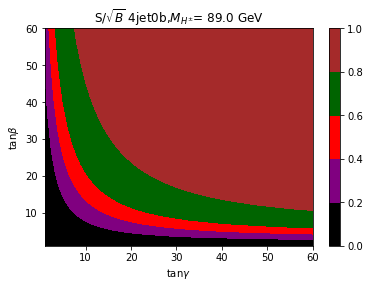}
    }\\
    \subfigure
    {
        \includegraphics[scale=.5]{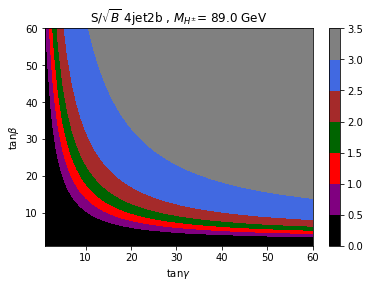}
    }
    \subfigure
    {
        \includegraphics[scale=.5]{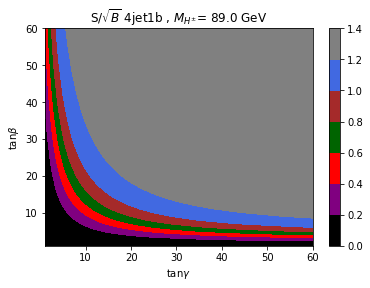}
    }
    \caption
    { The flipped 3HDM with $\theta = - \frac{\pi}{2.1}, \delta = 0$, and  $M_{H^{\pm}} = 89$ GeV. Top Panel: Significance ($S/\sqrt B$)
at a single LEP2 experiment in the 4-jet channel without $b$-tagging, in $[\tan\gamma, \tan\beta]$ plane. Left Bottom Panel: $S/\sqrt B$ 
in the 4-jet channel with two $b$-tags, in $[\tan\gamma, \tan\beta]$ plane. Right Bottom Panel: $S/\sqrt B$ in the 4-jet channel 
with one $b$-tag, in $[\tan\gamma, \tan\beta]$ plane
    } \label{4jallb89}
\end{figure}

\begin{figure}[h!]
    \centering
    \subfigure[]
    {
        \includegraphics[scale=.55]{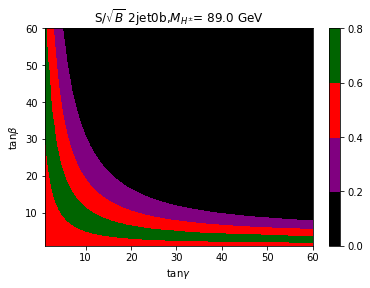}
    }
    \subfigure[]
    {
        \includegraphics[scale=.55]{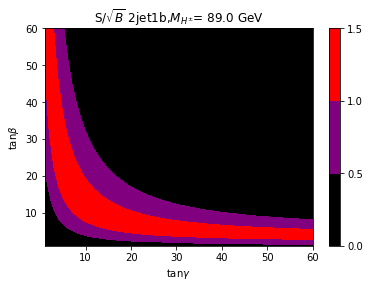}
    }
    \caption
    {The flipped 3HDM with $\theta = - \frac{\pi}{2.1}, \delta = 0$, and  $M_{H^{\pm}} = 89$ GeV. Left Panel: Significance ($S/\sqrt B$) 
at a single LEP2 experiment in the 2-jet channel without $b$-tagging, in $[\tan\gamma, \tan\beta]$ plane. 
Right Panel: $S/\sqrt B$ in the 2-jet channel with one $b$-tag, in $[\tan\gamma, \tan\beta]$ plane.}\label{2jallb89}
\end{figure}

The BRs of $H^\pm$ as functions of the four parameters 
($\tan\beta$, $\tan\gamma$, $\theta$, $\delta$) have been studied
in detail in \cite{Akeroyd:2018axd}, and the parameter space for a 
dominant BR$(H^\pm\to cb)>50\%$ was displayed. 
In Fig.~\ref{br80} (left panel) 
contours of BR($H^{\pm} \to cb$) are shown in the plane 
$[\tan\gamma, \tan\beta]$, for $M_{H^{\pm}}$ = 80 GeV (the 
results with $M_{H^{\pm}}$ = 89 GeV are essentially 
identical). We fix $\theta = -\frac{\pi}{2.1}$ and $\delta=0$, for which
a sizeable part of the plane $[\tan\gamma, \tan\beta]$ gives 
BR$(H^\pm\to cb)>60\%$, with around $80\%$ being the maximum value.
Similar plots (for different choices of $\theta$) can be found in 
\cite{Akeroyd:2018axd}.  This parameter choice for $\theta$ and $\delta$ 
will be used
in Fig.~\ref{br80} to Fig.~\ref{2jallb89}, with all these plots being 
shown in the plane $[\tan\gamma, \tan\beta]$.
In Fig.~\ref{br80} (right panel) contours of Re($XY^*$)
are shown, with the  
region $-1.1 \le {\rm Re}(XY^*)\le 0.7$ being (roughly) consistent with 
the limits on 
BR($b\to s\gamma$) for $M_{H^{\pm}} = 80$ GeV. Clearly the majority of the
plane $[\tan\gamma, \tan\beta]$ satisfies this constraint, and thus
the large values of BR$(H^\pm\to cb)$ in Fig.~\ref{br80} (left panel) 
are permissible. 

In Fig.~\ref{brcbcs80} (left panel) and Fig.~\ref{brcbcs80} 
(right panel) contours of BR$(H^\pm \to cs)$ and   
BR($H^{\pm} \to \tau \nu$) (respectively) are displayed. For this choice of 
$\theta = -\frac{\pi}{2.1}$ and $\delta=0$ one can see that  
BR$(H^\pm \to cs)\approx 35\%$ 
and BR($H^{\pm} \to \tau \nu)\approx 65\%$ when BR$(H^\pm\to cb)$
is small (corresponding to small $\tan\beta$ and $\tan\gamma$).
In Fig.~\ref{background80} the sums and products of BRs of $H^\pm$
are displayed, which will aid the understanding of 
the statistical significances that are displayed in 
Fig.~\ref{4j0b80} to Fig.~\ref{2jallb89}.
In Fig.~\ref{background80} (top left panel) contours of 
BR$(H^\pm \to hadrons)$ are shown, where ``$hadrons$'' refers to the sum of 
$cs$ and $cb$. In  Fig.~\ref{background80} 
(top right panel) and Fig.~\ref{background80} (bottom panel)
contours of BR$(H^\pm \to hadrons) \times 
{\rm BR}(H^\pm \to \tau\nu)$ and 
BR$(H^\pm \to cb) \times {\rm BR}(H^\pm \to \tau\nu)$ (respectively) are shown.
Note that BR$(H^\pm \to cb) \times {\rm BR}(H^\pm \to \tau\nu)$ is maximised
(taking a value of around 0.14) in a band that is away from the region
of both $\tan\beta$ and $\tan\gamma$ being small or large.

\begin{figure}[b]
    \centering
    \subfigure
    {
        \includegraphics[scale=.5]{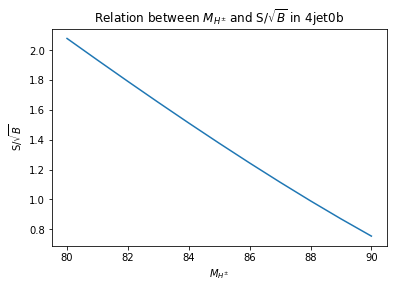}
    }\\
        \subfigure
    {
        \includegraphics[scale=.5]{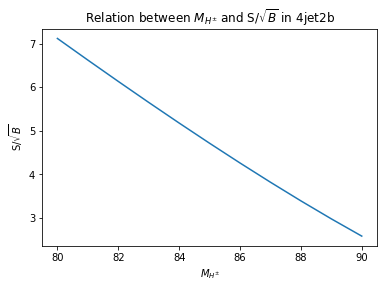}
    }
    \subfigure
    {
        \includegraphics[scale=.5]{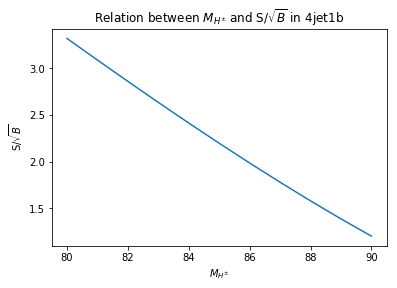}
    }
    \caption
    {Dependence of $S/\sqrt B$ on $M_{H^\pm}$, with BR$(H^\pm \to cb) = 0.8$ (near maximal) and  BR$(H^\pm \to cs) = 0.2$, at
a single LEP2 experiment.  Top Panel: 
In 4-jet channel without $b$-tagging. Left Bottom Panel: In 4-jet channel with two $b$-tags.  Right Bottom Panel: In 4-jet
channel with one $b$-tag. } \label{sig_mhch4j}
\end{figure}

\begin{figure}[t!]
    \centering
    \subfigure
    {
        \includegraphics[scale=.5]{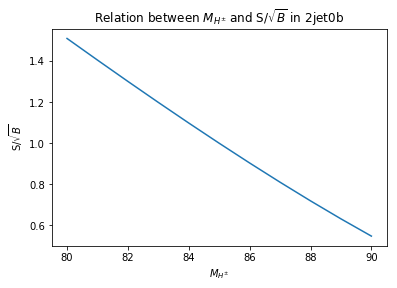}
    }
    \subfigure
    {
        \includegraphics[scale=.5]{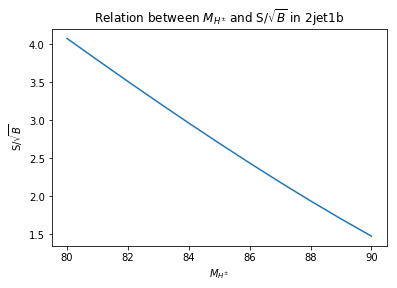}
    }
     \caption{Dependence of $S/\sqrt B$ on $M_{H^\pm}$, with BR$(H^\pm \to hadrons) = 0.5$ and  BR$(H^\pm \to \tau\nu) = 0.5$
at a single LEP2 experiment.  Left Panel: 
In 2-jet channel without $b$-tagging. Right Panel: In 2-jet channel with one $b$-tag. }  \label{sig_mhch2j}
\end{figure}

\begin{figure}[t!]
    \centering
    \subfigure
    {
        \includegraphics[scale=.5]{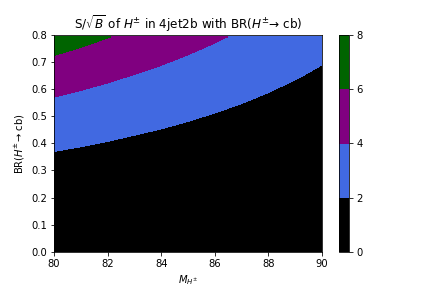}
    }
    \subfigure
    {
        \includegraphics[scale=.5]{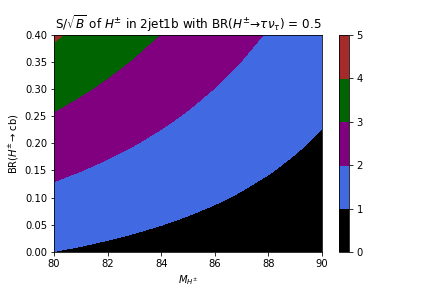}
    }
     \caption{Values of $S/\sqrt B$ in the plane $[M_{H^\pm}, {\rm BR}(H^\pm\to cb)]$ at a single LEP2 experiment.\\
Left Panel: In 4-jet channel with two $b$-tags, with BR$(H^\pm\to cb)$+BR$(H^\pm\to cs)=1$.
Right Panel: In 2-jet channel with one $b$-tag, with BR$(H^\pm \to \tau\nu) = 0.5$, and BR$(H^\pm\to cb)$+BR$(H^\pm\to cs)=0.5$.
 }  \label{BRcbplot}
\end{figure}

\begin{figure}[t!]
    \centering
    \subfigure
    {
        \includegraphics[scale=.5]{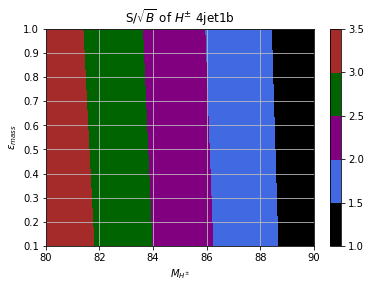}
    }
    \subfigure
    {
        \includegraphics[scale=.5]{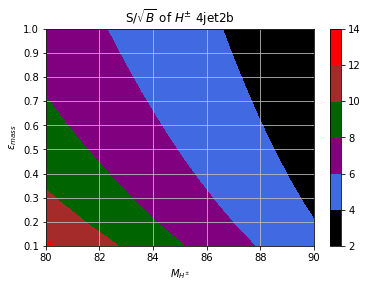}
    }
    \caption
    {Dependence of $S/\sqrt B$ on $M_{H^\pm}$ and on invariant mass cut ($\epsilon_{mass}$) at a single LEP experiment, 
with BR$(H^\pm \to cb) = 0.8$ and BR$(H^\pm \to cs) = 0.2$. Left Panel: 4-jet channel with one $b$-tag. Right Panel: 4-jet channel 
with two $b$-tags.}  \label{mass_mhch4j}
\end{figure}

In Fig.~\ref{4j0b80} to Fig.~\ref{2jallb89} the statistical significances $(S/\sqrt B)$ are
shown in the five channels (three with $b$-tagging and
two without $b$-tagging) at a single experiment at LEP2 in the plane $[\tan\gamma, \tan\beta]$,
for $M_{H^{\pm}}$ = 80 GeV and 89 GeV. In the three panels in Fig.~\ref{4j0b80}, 
$S_{4jnobtag}/\sqrt {B_{4jnobtag}}$ for the 4-jet channel (top panel), $S_{4j2btag}/\sqrt {B_{4j2btag}}$ 
(left bottom panel), and $S_{4j1btag}/\sqrt {B_{4j1btag}}$ (right bottom panel) are plotted, with $M_{H^{\pm}}$ = 80 GeV. 
For the case with no $b$-tagging (which corresponds to the experimental searches) one can see
that the largest signal satisfies $2 < S/\sqrt B< 2.5$ and arises in the region where BR$(H^\pm \to hadrons)$ 
in Fig.~\ref{background80} (top left panel) is largest. The maximum $S/\sqrt B$ is less than 
2.5 for this choice of $M_{H^{\pm}}$ = 80 GeV, and this is roughly consistent with the  OPAL limits that ruled
out $M_{H^{\pm}}< 80$ GeV for BR$(H^\pm \to hadrons)=100\%$. For the case with two $b$-tags it is evident that $S/\sqrt B$
can be greatly increased with respect to the case with no $b$-tag. A large part of the $[\tan\gamma, \tan\beta]$ plane has
$S/\sqrt B>3$, with $S/\sqrt B\approx 8$ being possible. Note that these significances are for a single LEP2 experiment, and 
thus a $3\sigma$ signal at all four experiments might approach the $5\sigma$ threshold for discovery if the
four searches are combined. The individual values
of $S$ and $B$ will be shown in tables and discussed later. For the case with one $b$-tag it is found that the 
values of $S/\sqrt B$ (at a given point in the plane)  are slightly
larger than those for the case with no $b$-tag. Although the
background in the one $b$-tag channel is smaller, the signal has decreased such that the ratio $S/\sqrt B$ does not improve greatly compared to the case with no $b$-tag.

In the two panels in Fig.~\ref{2jallb80}, $S_{2jnobtag}/\sqrt {B_{2jnobtag}}$ for the 2-jet channel (left panel) and $S_{2j1btag}/\sqrt {B_{2j1btag}}$ 
(right panel) are plotted, with $M_{H^{\pm}}$ = 80 GeV. For the case with no $b$-tagging one sees that the largest $S/\sqrt B$ is around
1.6, and arises in the region where BR$(H^\pm \to hadrons) \times {\rm BR}(H^\pm \to \tau\nu)$
in Fig.~\ref{background80} (top right panel) is largest. Again, this maximum value for $S/\sqrt B$ is roughly consistent with the OPAL limits that ruled
out $M_{H^{\pm}}< 80$ GeV in the 2-jet channel when BR$(H^\pm \to hadrons) \times {\rm BR}(H^\pm \to \tau\nu)$ is at its maximum 
value. For the case with one $b$-tag it is evident that $S/\sqrt B$
can be somewhat increased with respect to the case with no $b$-tag, but the gain is less than that in the 4-jet channel with two $b$-tags.
Values of $S/\sqrt B$ up to 3.2 can be obtained in the region in  Fig.~\ref{background80} (bottom panel) where BR$(H^\pm \to cb) \times {\rm BR}(H^\pm \to \tau\nu)$ is largest.

Fig.~\ref{4jallb89} and Fig.~\ref{2jallb89} are the same as Fig.~\ref{4j0b80} and Fig.~\ref{2jallb80} respectively but with 
$M_{H^\pm}=$ 89 GeV instead of 80 GeV. The maximum $S/\sqrt B$ has dropped by roughly a factor of 2 for the 4-jet channel with two $b$-tags and for the
2-jet channel with one $b$-tag. This decrease is due to the reduction in the cross-section for $e^+e^-\to H^+H^-$ when going from
$M_{H^\pm}=$ 80 GeV to $M_{H^\pm}=$ 89 GeV. As mentioned earlier, a $3\sigma$ signal at each LEP2 experiment 
might become close to $5\sigma$ evidence by combining all four experiments. Hence a discovery for $M_{H^\pm}=$ 89 GeV is possible
in the most optimistic scenario of BR($H^\pm\to cb$) close to $80\%$.

In Fig.~\ref{sig_mhch4j} the dependence of $S/\sqrt B$ on $M_{H^\pm}$ is shown for the 4-jet channel, 
fixing BR$(H^\pm \to cb) = 0.8$ (i.e. near maximal) and BR$(H^\pm \to cs) = 0.2$.
The top panel, left bottom panel and right bottom panel are for the channels without $b$-tagging, two $b$-tags, and 
one $b$-tag respectively. One can see that the dependence is roughly linear, and that a $5\sigma$ signal at a single
LEP2 experiment is possible in the 4-jet channel with two $b$-tags up to around $M_{H^\pm}=84$ GeV.

In Fig.~\ref{sig_mhch2j} the dependence of $S/\sqrt B$ on $M_{H^\pm}$ is shown for the 2-jet channel, 
with BR$(H^\pm \to cs) = 0.1$, BR$(H^\pm \to cb) = 0.4$, and BR$(H^\pm \to \tau\nu) = 0.5$ (i.e. close to the optimum scenario
for discovery in this channel).
Note that this choice of BR$(H^\pm \to cb)\times$BR$(H^\pm \to \tau\nu) = 0.2$ is used for illustration, and is larger than the 
maximum value of this product in Fig.~\ref{background80} (bottom panel) with $\theta = -\pi/2.1$ and $\delta = 0$. Again, one sees a roughly linear dependence on $M_{H^\pm}$.
 In Fig.~\ref{BRcbplot}, $S/\sqrt B$ is plotted in the plane $[M_{H^\pm}, {\rm BR}(H^\pm\to cb)]$. In the
left panel we show the results in the 4-jet channel with two $b$-tags, with BR$(H^\pm\to cb)$+BR$(H^\pm\to cs)=1$.
It can be seen that  BR$(H^\pm\to cb)>0.4$ is required in order to obtain $S/\sqrt B>2$ for
$M_{H^\pm}=80$ GeV at a single experiment.
In the right panel we show the results for the 
2-jet channel with one $b$-tag, taking BR$(H^\pm \to \tau\nu) = 0.5$, and BR$(H^\pm\to cb)$+BR$(H^\pm\to cs)=0.5$.
It can be seen that  BR$(H^\pm\to cb)>0.15$ is required in order to obtain $S/\sqrt B>2$.

It is clear from the above plots that the 4-jet channel with two $b$-tags offers the largest values of $S/\sqrt B$.
In Tab.~\ref{tab:sigtable4j} the individual values of $S$ and $B$ (and $S/\sqrt B$) are shown for $M_{H^\pm}=80$ GeV, 85 GeV and 89 GeV
in 4-jet channels,  with BR$(H^\pm \to cb) = 0.8$ and  BR$(H^\pm \to cs) = 0.2$. It can be seen that the background decreases significantly as
each $b$-tag is applied, and there are still a significant number of events ($S\approx 9$) in the 4-jet channel with 2 $b$-tags 
for $M_{H^\pm}=89$ GeV. Around 7 of 9 background events in the $4j2b$ channel are from the two-fermion background, and an invariant mass
cut could further reduce this background (see later). 

In Tab.~\ref{tab:sigtable2j} the individual values of $S$ and $B$ (and $S/\sqrt B$) are shown for $M_{H^\pm}=80$ GeV, 85 GeV and 89 GeV
in 2-jet channels, with  BR$(H^{\pm} \to cb) =  0.4$, BR$(H^{\pm} \to cs) =  0.1$ and BR$(H^{\pm} \to \tau\nu) =  0.5$. Again, the
background has decreased significantly with the $b$-tag, and there are still a reasonable number of events ($S\approx 6$) in the 2-jet channel with a $b$-tag 
for $M_{H^\pm}=89$ GeV.

\begin{table} [!b]
\begin{center}
\begin{tabular}{|c||c|c|c|c|c|c|c|}
\hline
$M_{H^{\pm}}$
& 80 GeV & 85 GeV &  89 GeV &  80 GeV & 85 GeV &  89 GeV &  {}\\ \hline
& { $S$ } & {$S$} &  {$S$} &  { $\frac{S}{\sqrt{B}}$} &  { $\frac{S}{\sqrt{B}}$} &  { $\frac{S}{\sqrt{B}}$} &  { $B$}\\ \hline
4j0b 
&  {\color{black} 69.50} & {\color{black} 46.01} & {\color{black} 29.07}  & {\color{black} 2.08} & {\color{black} 1.38} & {\color{black} 0.87} & {\color{black} 1117.8}\\
4j1b 
& {\color{black} 31.74} & {\color{black} 21.01} & {\color{black}  13.27 }  & {\color{black} 3.32}  & {\color{black} 2.20}  & {\color{black} 1.39} & {\color{black}  91.44}\\
4j2b 
& {\color{black} 22.43} & {\color{black}14.85} & {\color{black} 9.38} & {\color{black} 7.12}  & {\color{black} 4.71}  & {\color{black}3.00} & {\color{black} 9.94} \\
\hline
\end{tabular}
\end{center}
\caption{Number of signal events ($S$), number of background events ($B$), and corresponding significances ($\frac{S}{\sqrt{B}}$) in 4-jet channels at 
single experiment at LEP2.
Results are shown for $M_{H^{\pm}}$ = 80, 85, 89 GeV, with BR$(H^{\pm} \to cb) =  0.8$ and BR$(H^{\pm} \to cs) =  0.2$.}\label{tab:sigtable4j}
\end{table}

\begin{table} [!b]
\begin{center}
\begin{tabular}{|c||c|c|c|c|c|c|c|}
\hline
$M_{H^{\pm}}$ 
& {80 GeV} & {85 GeV} &  {89 GeV} &  {80 GeV} &  {85 GeV} &  {89 GeV} &  {}\\ \hline
& {$S$} & {$S$} &  {$S$} &  {$\frac{S}{\sqrt{B}}$} &  {$\frac{S}{\sqrt{B}}$} &  {$\frac{S}{\sqrt{B}}$} &  {$B$}\\ \hline
2j0b 
&  {\color{black} 26.89} & {\color{black} 17.80} & {\color{black} 11.24}  & {\color{black} 1.51} & {\color{black} 1.00} & {\color{black} 0.63} & {\color{black} 316.9}\\
2j1b 
& {\color{black} 15.28} & {\color{black} 10.11} & {\color{black} 6.39 }  & {\color{black} 4.08}  & {\color{black} 2.70}  & {\color{black} 1.71} & {\color{black}  14.04}\\
\hline
\end{tabular}
\end{center}
\caption{Number of signal events ($S$), number of background events ($B$), and corresponding significances ($\frac{S}{\sqrt{B}}$) in 2-jet channels at a single
experiment at LEP2.
Results are shown for $M_{H^{\pm}}$ = 80, 85, 89 GeV, with BR$(H^{\pm} \to cb) =  0.4$, BR$(H^{\pm} \to cs) =  0.1$ and BR$(H^{\pm} \to \tau\nu) =  0.5$. }\label{tab:sigtable2j}
\end{table}

As mentioned above in the discussion of Tab.~\ref{tab:sigtable4j}, the 2-fermion background accounts for most of the background in the $4j2b$ channel.
The invariant mass ($m_{jj}$) of two of the four jets from the 2-fermion background has a flat distribution (as can be seen in 
the OPAL search in \cite{Abbiendi:2008aa}), while 
the signal is mainly contained in the region of $m_{jj}$ between 80 GeV and 89 GeV. Hence we suggest that an 
invariant mass cut
which only keeps jets satisfying $80\,{\rm GeV}\, < m_{jj} < 89$ GeV could further improve $S/\sqrt B$ in the  $4j2b$ channel.
From a figure in \cite{Abbiendi:2008aa} we estimate that such a cut could reduce the 2-fermion background by a factor of 2, while preserving
the majority of the signal events of an $H^\pm$ with a mass between 80 GeV and 89 GeV. In Fig.~\ref{mass_mhch4j} the effect of the invariant mass cut efficiency ($\epsilon_{mass}$)
on $S\sqrt B$ in the 4-jet channel with one and two $b$-tags is shown. For illustration we vary $\epsilon_{mass}$ from 1 (i.e. no cut) to 0.1, with values
of $0.4 < \epsilon_{mass}<0.5$ being suggested by a figure in \cite{Abbiendi:2008aa}. For simplicity we assume that the signal is not affected by the invariant mass cut.
Taking  $\epsilon_{mass}=0.4$ and  $M_{H^\pm}=80$ GeV one can see from the right panel (for the two $b$-tag channel) that $S/\sqrt B$ improves from around 7 
($\epsilon_{mass}=1$) to 9 ($\epsilon_{mass}=0.4$).

Finally, we comment on a slight excess of events of greater than $2\sigma$ significance
that is present in the LEP working group combination of the searches for
$e^+e^-\to H^+H^-$ at all four experiments \cite{Abbiendi:2013hk}. The excess
occurs around $M_{H^\pm}=89$ GeV, BR$(H^\pm\to hadrons)=65\%$ and BR$(H^\pm\to \tau\nu)=35\%$,
and in our earlier work \cite{Akeroyd:2016ssd,Akeroyd:2018axd} we suggested the possibility
of this being due to an $H^\pm$ of a 3HDM. If such an excess is genuine, and if
a large fraction of the hadronic BR is from $H^\pm\to cb$ decays, then 
$b$-tagging would increase the significance. In Tab.~\ref{tab:sigtable}
we show the values of $S$, $B$ and $S/\sqrt B$ for  $M_{H^\pm}=88$ GeV, 89 GeV
and 90 GeV. We take BR$(H^\pm\to cb)=50\%$ and BR$(H^\pm\to cs)=15\%$ (in order to
obtain BR$(H^\pm\to hadrons)=65\%$), and fix BR$(H^\pm\to \tau\nu)=35\%$.
From Tab.~\ref{tab:sigtable} it can be seen that the $4j0b$ and $2j0b$ channels (i.e. the current searches)
give significances of 0.37 and 0.57 respectively for $M_{H^\pm}=89$ GeV. These numbers are
for a single LEP2 experiment, and so it is conceivable that the 
combination of four experiments
could give the observed $2\sigma$ excess, especially if there has been
an upward fluctuation. In the $4j2b$ and $2j1b$
channels these significances increase to 1.17 and 1.51 respectively (i.e. a factor
of three improvement), with the 
number of signal events ($S$) still being above three events in each channel.
Consequently, if the excess is genuine then its significance could be 
significantly increased in the $4j2b$ and $2j1b$
channels, assuming that BR$(H^\pm\to cb)$ is large.
As discussed in \cite{Akeroyd:2018axd} such a signal might also 
show up at the LHC in the channel $t\to H^\pm b$, 
which currently has sensitivity to the region 80 GeV $< M_{H^\pm}< 90$ GeV for
$H^\pm\to \tau\nu$ decays (but does not yet have sensitivity to $H^\pm\to hadrons$ decays in this mass region).
However, if the couplings $|X|$ and $|Y|$ (which determine BR($t\to H^\pm b$)) are
sufficiently small then such an $H^\pm$ would remain hidden from LHC searches.
At $e^+e^-$ colliders the production channel $e^+e^-\to H^+H^-$ does not
depend on $|X|$ and $|Y|$, and a high luminosity $e^+e^-$ collider would be able 
to probe the region  80 GeV $< M_{H^\pm}< 90$ GeV irrespective of  $|X|$ and $|Y|$.

\begin{table} [h]
\begin{center}
\begin{tabular}{|c||c|c|c|c|c|c|c|}
\hline
$M_{H^{\pm}}$ 
&  88 GeV & 89 GeV &   90 GeV &  88 GeV &  89 GeV &  90 GeV &  {}\\ \hline
& { $S$ } & {$S$} &  {$S$} &  { $\frac{S}{\sqrt{B}}$} &  { $\frac{S}{\sqrt{B}}$} &  { $\frac{S}{\sqrt{B}}$} &  {$B$}\\ \hline
4j0b 
&  {\color{black} 13.98} & {\color{black} 12.28} & {\color{black} 10.64}  & {\color{black} 0.42} & {\color{black} 0.37} & {\color{black} 0.32} & {\color{black} 1117.8}\\
4j1b 
& {\color{black} 6.47} & {\color{black} 5.68} & {\color{black} 4.93 }  & {\color{black} 0.68}  & {\color{black} 0.59}  & {\color{black} 0.52} & {\color{black}  91.44}\\
4j2b 
& {\color{black} 4.21} & {\color{black}3.7} & {\color{black} 3.21} & {\color{black} 1.34}  & {\color{black} 1.17}  & {\color{black}1.02} & {\color{black} 9.94} \\
2j0b 
& {\color{black}11.65} & {\color{black} 10.23} & {\color{black} 8.87} & {\color{black}  0.65}  & {\color{black} 0.57}  & {\color{black}  0.5} & {\color{black} 316.9}\\
2j1b
& {\color{black} 6.43} & {\color{black} 5.65} & {\color{black}4.89}& {\color{black} 1.72} & {\color{black} 1.51}  & {\color{black} 1.31} & {\color{black} 14.04} \\
\hline
\end{tabular}
\end{center}
\caption{Number of signal events ($S$), number of background events ($B$), and corresponding significances ($\frac{S}{\sqrt{B}}$) in five channels at a single
experiment at LEP2. Results
are shown for $M_{H^{\pm}} = 88,\, 89, \,90$ GeV, with BR$(H^{\pm} \to cb) =  0.5$, BR$(H^{\pm} \to cs) =  0.15$, and BR$(H^{\pm} \to \tau \nu) = 0.35$. }\label{tab:sigtable}
\end{table}

\subsection{Prospects for detecting $H^\pm\to cb$ at CEPC/FCC-ee}

Future $e^+e^-$ colliders \cite{deBlas:2019rxi} are being discussed, which would offer precise measurements of the properties of the 125 GeV neutral Higgs boson.
Such colliders would also permit detailed studies of a light charged Higgs boson. There are two proposals for a circular
$e^+e^-$ collider with a period of operation at $\sqrt{s} = 240$ GeV: CEPC in China \cite{CEPCStudyGroup:2018ghi} and FCC-ee \cite{Abada:2019lih} 
at CERN. These colliders would produce a large
number of $H^+H^-$ events with a mass of up to $M_{H^\pm}=120$ GeV. The integrated luminosity at this energy
is expected to be
of the order of a few ${\rm ab}^{-1}$, which is roughly a thousand times larger than the total integrated luminosity
taken at a single LEP2 experiment (0.6 ${\rm fb}^{-1}$). Two linear $e^+e^-$ colliders are also being discussed, the International
Linear Collider (ILC) \cite{Bambade:2019fyw} and the Compact Linear Collider (CLIC) \cite{deBlas:2018mhx}, which will both offer the possibility of energies much higher than
 $\sqrt{s} = 240$ GeV. In this work we will consider the detection prospects of the decay channel $H^\pm\to cb$ at 
 $\sqrt{s} = 240$ GeV only. As mentioned earlier, BR$(H^\pm\to cb)$ is
expected to be at most of the order of $1\%$ in the Type I, Type II, and leptonic specific 3HDMs.
Only the flipped and democratic 3HDMs can have BR$(H^\pm\to cb)$ significantly 
larger than $1\%$. Consequently, precise measurements of  BR$(H^\pm\to cb)$ could shed light on which 3HDM Yukawa structure is realised. 
It is our aim to see if CEPC/FCC-ee would have sensitivity to smaller values (of the order of a few percent) for BR$(H^\pm\to cb)$. For
the number of background events we use the values from LEP2 (for which $\sqrt s \approx 200$ GeV) 
for simplicity.
The parameter $\epsilon_c$ was taken to be 0.06 in our analysis at LEP2. At CEPC/FCC-ee we expect that this efficiency would be
improved, and thus we vary it in the range $0.01 < \epsilon_c < 0.06$. We keep $\epsilon_b$ and $\epsilon_j$ at the LEP2 values.
The input parameters for the study of the detection prospects of $H^\pm\to cb$ at $\sqrt s=240$ GeV are summarised in 
Tab.~\ref{tab:collider}.

\begin{figure}[t!]
    \centering
    \subfigure
    {
        \includegraphics[scale=.5]{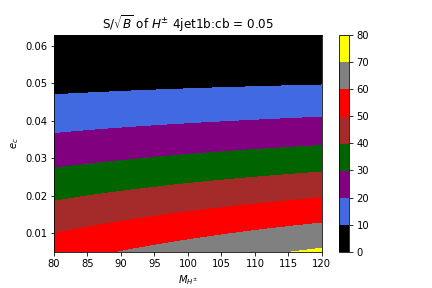}
    }
    \subfigure
    {
        \includegraphics[scale=.5]{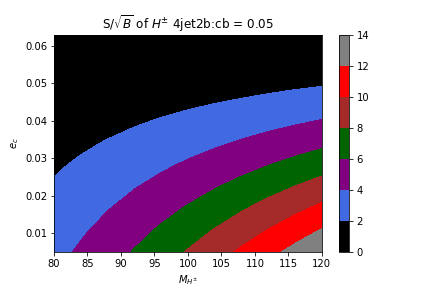}
    }
    \caption
    {Dependence of $S/\sqrt B$ on $M_{H^\pm}$ and $\epsilon_c$ at $\sqrt s=240$ GeV (CEPC/FCC-ee), with BR$(H^\pm \to cb) = 0.05$
and BR$(H^\pm \to cs) = 0.95$. Left Panel: 4-jet channel with one $b$-tag. Right Panel: 4-jet channel with two $b$-tags.}  
\label{ec_mhch4jcepc}
\end{figure}

\begin{figure}[t!]
    \centering
    \subfigure
    {
        \includegraphics[scale=.5]{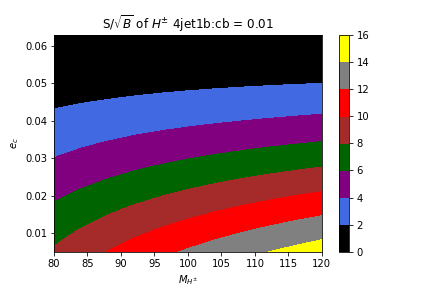}
    }
    \subfigure
    {
        \includegraphics[scale=.5]{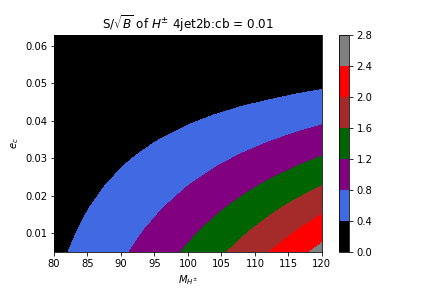}
    }\\
     \subfigure
  {
         \includegraphics[scale=0.5]{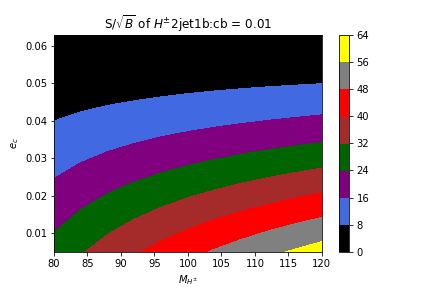}
    }
    \caption
       { Dependence of $S/\sqrt B$ on $M_{H^\pm}$ and $\epsilon_c$ at $\sqrt s=240$ GeV (CEPC/FCC-ee), with
BR$(H^\pm \to cb) = 0.01$, BR$(H^\pm \to cs) = 0.50$, and BR$(H^\pm \to \tau\nu) = 0.49$.
 Top Left Panel:  4-jet channel with one $b$-tag. Top Right Panel: 4-jet channel with two $b$-tags.   Bottom Panel: 2-jet channel 
with one b-tag.}  \label{ec_mhch4jcb0.01cepc}
\end{figure}

In Fig.~\ref{ec_mhch4jcepc} the dependence of $S/\sqrt B$ on $M_{H^\pm}$ and $\epsilon_c$ at $\sqrt s=240$ GeV is shown, 
with BR$(H^\pm \to cb) = 0.05$ (which would only be possible in flipped/democratic 3HDMs)
and BR$(H^\pm \to cs) = 0.95$ (i.e. the decays to leptons are absent). 
The left panel is for the 4-jet channel with one $b$-tag
and the right panel is for the 4-jet channel with two $b$-tags. In the 4-jet channel with one $b$-tag it can be
seen that very large values of $S/\sqrt B$ can be achieved (e.g. $S/\sqrt B \approx 30$ for $\epsilon_c=0.03$ and
$M_{H^\pm}=90$ GeV), and thus precise measurements of BR$(H^\pm \to cb)$ would be obtained over a wide
region of the plane $[M_{H^\pm}, \epsilon_c]$.
 Note that the values
of $S/\sqrt B$ are much lower in the 4-jet channel with two $b$-tags (right panel). This is because BR$(H^\pm \to cb) = 0.05$, leading 
to a reduced number of signal events with two $b$-quarks compared to the case at LEP2 where the optimum 
scenario of BR$(H^\pm \to cb) = 0.8$ was considered.

In Fig. \ref{ec_mhch4jcb0.01cepc}, the dependence of $S/\sqrt B$ on $M_{H^\pm}$ and $\epsilon_c$ at $\sqrt s=240$ GeV is shown,
with BR$(H^\pm \to cb) = 0.01$, BR$(H^\pm \to cs) = 0.50$, and BR$(H^\pm \to \tau\nu) = 0.49$.
The top left panel is for the 4-jet channel with one $b$-tag, the top right panel is for the
4-jet channel with two $b$-tags, and the bottom panel is for the 2-jet channel with one $b$-tag.
In the 4-jet channel with one $b$-tag and the
 2-jet channel with one $b$-tag a clear signal (and hence a precise measurement) can be achieved over a wide
region of the plane $[M_{H^\pm}, \epsilon_c]$. This would
establish the presence of $H^\pm\to cb$ decays even for BR$(H^\pm \to cb) = 0.01$, 
a BR that is theoretically possible in all five Yukawa structures.

\section{Conclusions}  

The decay channel $H^{\pm} \to c b$ can have a large BR (up to 80$\%$) in the flipped and democratic 3HDMs for $M_{H^{\pm}} < M_t$, and be compatible with constraints from $b \to s \gamma$. The current search at the LHC (with $\sqrt{s} = 8$ TeV, ${\cal L} = 20$ fb$^{-1}$) for $t \to H^{\pm}b$ followed by $H^{\pm} \to cb$ is not sensitive to the region 80 GeV $\leq M_{H^{\pm}} \leq$ 90 GeV, although sensitivity might be reached in future searches. LEP2 searched for $e^{+} e^- \to H^{+} H^-$, assuming the main
decay channels to be $H^{\pm} \to  hadrons$ and $H^{\pm} \to \tau\nu$. In the region 80 GeV $\leq M_{H^{\pm}} \leq$ 90 GeV, a sizeable part of the plane [BR($H^{\pm} \to  hadrons$), $M_{H^{\pm}}$] is not excluded at LEP2 if
BR($H^{\pm} \to  hadrons$) is dominant. If BR($H^{\pm} \to  cb$) were large then more of the region 80 GeV $\leq M_{H^{\pm}} \leq$ 90 GeV could be probed at LEP2 by adding one or more $b$-tags to the existing search strategy. We evaluated the significances ($S/\sqrt{B}$) for $H^{\pm} \to  cb$ decays in
three channels by taking the selection efficiencies and backgrounds from the OPAL searches, and applying realistic $b$-tagging and fake $b$-tagging efficiencies. In the optimum scenario of BR($H^{\pm} \to  cb$) = 80\% (BR($H^{\pm} \to  cb$) = 40\% for 2-jets), it was shown that $S/\sqrt{B}$ as large as 7, 3 and 4 could be obtained
for $M_{H^{\pm}}$ = 80 GeV in the three channels i) 4-jet plus two b-tags, ii) 4-jet plus one btag, and iii) 2-jets plus one b-tag, respectively. These significances decrease to roughly 3, 1.4 and 1.7 respectively for $M_{H^{\pm}}$ = 89 GeV, but would be increased by combining all four
experiments. Consequently, LEP2 has the capability to exclude or discover a $H^{\pm}$ with a large BR($H^{\pm} \to  cb$), and with a mass in the region 80 GeV $\leq M_{H^{\pm}} \leq$ 90 GeV. We commented
on a $> 2\sigma$ excess at around $M_{H^{\pm}}$ = 89 GeV and BR($H^{\pm} \to  hadrons$) $\approx 65\%$ in the LEP working group combination. Under the assumption that such an excess is genuine and has
a large BR($H^{\pm} \to  cb$), it was shown that its significance could be increased significantly in two of the three channels with $b$-tagging. We encourage an updated LEP2 search for $H^{\pm}$ that includes $b$-tagging as suggested above. This would become especially important if the
LHC eventually obtains evidence for an $H^{\pm}$ with 80 GeV $\leq M_{H^{\pm}} \leq$ 90 GeV and a large BR($H^{\pm} \to  cb$).

In contrast to hadron colliders, the cross-section for $H^{\pm}$ at LEP2 does not depend on the magnitude of the Yukawa couplings. Hence a light $H^{\pm}$ with small Yukawa couplings could escape detection at the LHC, but be discovered at LEP2 or at future $e^+ e^-$ colliders. Even if a light $H^{\pm}$ is discovered at the LHC, future $e^+ e^-$ colliders would be able to measure its BRs much more precisely in order to shed light on the underlying Higgs structure. We evaluated $S/\sqrt{B}$ for $H^{\pm} \to  cb$ decays at a proposed $e^+ e^-$ collider (CEPC/FCC-ee) of
$\sqrt{s} = 240$ GeV and found that BR($H^{\pm} \to  cb$) = 1\% (which is possible in all 2HDMs/3HDMs) would give a clear signal. In the context of 3HDMs, the flipped and democratic structures are the only ones which can have BR($H^{\pm} \to  cb$) significantly greater than 1\%, and so precise
measurements of this channel could provide evidence for these models.

\section*{Acknowledgements}
SM is supported in part through the NExT Institute and the STFC Consolidated Grant ST/L000296/1. SM and MS acknowledge the H2020-MSCA-RISE-2014 grant no. 645722 (NonMinimalHiggs). MS thanks Prof. Shinya Kanemura and Osaka University as well as Prof. Tetsuo Shindou and Kogakuin University for hospitality where parts of this work were carried out.


\end{document}